# Heat Engines running upon a Non-Ideal Fluid Model with Higher Efficiencies than upon the Ideal Gas Model

Abhimanyu S Madakavil[1,2], Ilki Kim[1]*

[1]Center for Energy Research and Technology, North Carolina A&T State University, Greensboro, NC 27411
[2]Department of Computational Science and Engineering, North Carolina A&T State University, Greensboro, NC 27411
E-mail: *hannibal.ikim@gmail.com

**Abstract**

We consider both Otto and Diesel heat engine cycles running upon the working substances modeled by the van der Waals fluid as a simple non-ideal gas model. We extensively perform the efficiency study of these model engines. Then we find that this "real" engine model can go beyond its ideal-gas counterpart in efficiency, whereas as well-known, the maximum Carnot efficiency is the same for both ideal and non-ideal gas engines. In fact, the higher the density of non-ideality is, the higher efficiency tends to be found, especially in the low-temperature regime, but with more shrinkage in the range of physically allowed compression ratio, determined by the Carnot upper bound, namely, the Second Law of thermodynamics. Our findings imply that in addition to the engine architectures and bath temperatures, the properties of working substances should also be taken into consideration in the performance study of heat engines, the theoretical model aspects of which have not sufficiently been discussed so far. It is straightforward that our methodology may also apply for other non-ideal fluid models.

## 1. Introduction

Heat engine is a central subject of any thermodynamics course in physics, physical chemistry, biology, and various engineering fields. Its discussion typically begins with the celebrated Carnot cycle consisting of two isothermal and two adiabatic (i.e., isentropic) processes and running upon the working substance described by the ideal gas model (satisfying $pv = RT$ for a mole of the gas). Due to its mathematical simplicity, this idealistic engine model is then straightforwardly employed for a discussion of the Second Law of thermodynamics. The Carnot efficiency is given by $\eta_{\mathrm{c}} = 1 - 1/t$, where the temperature ratio $t = T_h/T_c$ for hot and cold bath temperatures $T_h$ and $T_c$. In fact, this efficiency is true, independently of the engine size and regardless of the working fluid, i.e., even for an arbitrary non-ideal gas [1]. However, the behavior of efficiency with such independency and regardlessness may, in general, not be found for other cycles, such as the Otto and Diesel cycles to be in detailed consideration below.

As a simple representative of non-ideal fluid, the van der Waals (vdW) model has been widely used for an illustration of general behaviors of such a fluid [2], explicitly given by

$$p = \frac{RT}{v-b} - \frac{a}{v^2}, \qquad (1)$$

where the parameter $a > 0$ denotes a measure of the average attractive potential energy between the fluid particles, and the parameter $b$ is an effective molar volume of the fluid. In the dilute limit ($a, b \to 0$), this fluid model will reduce to the ideal gas. Various thermodynamic properties of this model, such as free energies, heat capacities and phase diagrams, have been extensively studied in [3,4]. On the other hand, an explicit and systematic analysis of the vdW fluid as a working substance of heat engine has not sufficiently been carried out yet, except for the cases of Carnot-type cycles (e.g., [5]).

The prototype of a real engine commonly used for vehicles is the reciprocating internal combustion engine. As its simple models, the Otto and the Diesel cycles are typically in consideration; the first cycle consists of two isochoric and two adiabatic processes while for the second one, an isobaric and an isochoric process as well as two adiabatic processes are given. It is then well-known from the undergraduate courses of engineering thermodynamics [6] that for a given compression ratio, the engine efficiencies of these two cycles running upon the ideal gas and driven reversibly are, each, smaller than the Carnot efficiency.

Furthermore, thermodynamics of small objects at the molecular scale in the low-temperature regime has recently attracted considerable interest, especially with an increase of the demand for optimal heat management in small and lightweight devices with high-level functionalities in modern electronics [7-9]. In this context, the (classical) vdW model may, by construction, reflect some molecular effects which will show up significantly in small engines with $1 \gg v \sim b$ (cf. Fig. 1). Therefore, it is also insightful to consider the schematic models of small internal combustion engines within the vdW fluid, i.e., without resort to quantum calculations, to observe a qualitative transition in efficiency with a decrease in engine size.

Here we will carry out the efficiency study of Otto and Diesel cycles running upon the vdW fluid, and compare their behaviors with that of the ideal-gas model, especially in the regime of small engine size and low temperature. The general layout of this paper is the following. In Sect. 2 we will briefly review the general frameworks of vdW fluid to be needed for our later discussions. In Sect. 3 we will derive exact expressions of the vdW Otto-cycle efficiency, with the help of a temperature-dependent heat capacity, and then discuss the results and their relations to actual real engines. In Sect. 4 the same analysis will take place for the Diesel cycle. Finally we will give the concluding remarks of this paper in Sect. 5.

## 2. Basics of van der Waals fluid

We begin with the expression of internal energy given by

$$dU = C_V dT + \left\{ T \left(\frac{\partial p}{\partial T}\right)_V - p \right\} dV , \quad (2)$$

which is true for an arbitrary gas equation of state [2]. Here the symbol $C_V$ denotes the heat capacity at constant volume. From the fact that Eq. (2) is an exact differential, it is required that

$$\left(\frac{\partial C_V}{\partial V}\right)_T = T \left(\frac{\partial^2 p}{\partial T^2}\right)_V , \quad (3)$$

which, for the vdW gas given by (1), easily reduces to

$$\left(\frac{\partial C_V}{\partial V}\right)_T = 0 . \quad (4)$$

Therefore, we see that the heat capacity of vdW gas is not a function of volume.

As well-known, differing from the ideal gas model, a real gas experiences the phase transition with a decrease of temperature (precisely speaking, only below the so-called critical temperature $T_{cr}$), typically first to the condensation phase, being a coexisting (i.e., two-phase) state of gas and liquid, and then continuing to its pure liquid phase with a further compression of volume [2]. In the vdW model, the temperature $T_{cr}$, as well as the critical pressure $p_c$ and critical volume $v_c$, can be evaluated exactly by rewriting Eq. (1) as a cubic equation in volume and then requiring both first and second derivatives of this cubic equation with respect to volume to vanish, since the critical point $(T_{cr}, p_c, v_c)$ has a flat inflection of isotherm curve in the p-v diagram, as depicted in Fig. 1 (the top curve therein). Then, it follows that $T_{cr} = 8a/(27Rb)$, $v_c = 3b$, and $p_c = a/(27b^2)$, expressed in terms of the vdW parameters [2].

In the two-phase state region of a typical real gas, pressure $p$ does not depend on volume $v$, as shown in Fig. 1 (see, e.g., [2] for the detail). Obviously, the vdW model cannot provide this behavior. Therefore, the two-phase state region should be treated separately. In fact, it has been shown in [3] that within the vdW model, (i) the molar volume $v_{GL}$ of two-phase region are required to satisfy the condition

$$\frac{1}{3} v_c + \frac{2\sqrt{T_r}}{9\sqrt{3}} v_c < v_{GL} < \left(\frac{9}{4T_r}\right) v_c , \quad (5)$$

where the reduced (dimensionless) temperature $T_r = T/T_{cr}$; (ii) accordingly, if the volume $v = v_G > \{9/(4T_r)\}v_c = 2a/(RT)$, then a substance with molar volume $v_G$ is in the gas phase; (iii) if $0 < v_L - v_c/3 < \{2\sqrt{T_r}/(9\sqrt{3})\}v_c$, equivalently, $0 < (v_L/b) - 1 < \{RbT/(2a)\}^{1/2}$, then a substance with molar volume $v_L$ is in the liquid phase. For the later purpose, we will below discuss heat capacities of a vdW fluid in those three different phases, respectively.

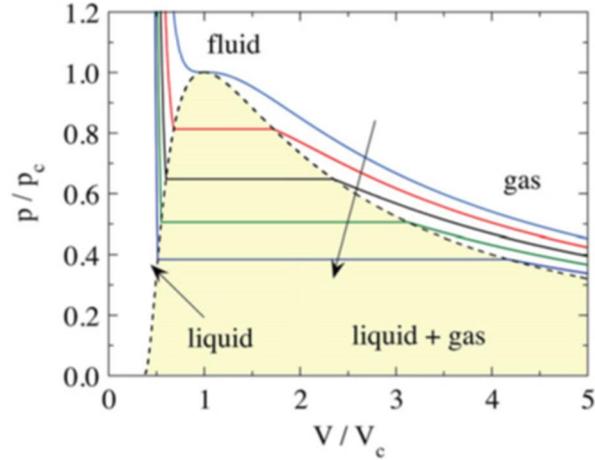

*Figure 1*. (Color online) *The isotherm curves in the diagram of (dimensionless) pressure $p/p_c$ versus (dimensionless) volume $v/v_c$ within the vdW fluid model, borrowed from [4]. From top to bottom of the curves: $T/T_{cr}$ = 1, 0.95, 0.9, 0.85, and 0.8. All white-colored region covers the single-phase states (i.e., gas, liquid, and undifferentiated fluid), fully described by Eq. (1), while the yellow-shaded region, on the other hand, the gas-liquid two-phase states, described by $(\partial p/\partial v)_T = 0$ with Eq. (5).*

First, the vdW molar heat capacity of pure gas phase may be given by $c_v = (d/2) R$ with $d$ being the number of degrees of freedom. It is the same as the heat capacity of ideal gas. Second, in the pure liquid phase, its vdW heat capacity is shown to be the same as that of pure gas phase, too [4]. However, it is easy to observe that this (classical) value $c_v$ of liquid in the low-temperature regime contracts the Third Law of thermodynamics saying that $c_v \to 0$ if $T \to 0$ [3]. In fact, the data from experiments demonstrate that the heat capacity of a liquid may differ from that of its gas phase [2]. To conform with the Third Law, we are now encouraged to introduce a temperature-dependent expression of the heat capacity model for the liquid phase,

$$c_v(T) = \frac{d}{2} \frac{RT}{(T+T_0)} \quad (6)$$

with the constant $T_0 > 0$. With $T_0 \to 0$ or in the high temperature regime $(T \to \infty)$, Eq. (6) obviously reduces to the ideal-gas value. Also, at an arbitrary temperature, this is smaller than the constant $(d/2)R$ being the gas-phase heat capacity. This aspect could be supported by the consideration that the translational and rotational motions are relatively frozen in a liquid [3,10]. Further, this model expression obviously differs from the heat capacity of the Einstein-Debye model being valid for the solid phase [2]. Eq. (6) will be employed below for our discussions of the following sections, in particular for the study of low-temperature effects.

Third, the heat capacity of gas-liquid mixture in the vdW model is shown to be temperature-dependent, as intensively studied in [4]. On the other hand, the prototype of internal combustion engine deals with the working fluid in a single phase [11]. Therefore, we will restrict our discussions below into the working substance in either pure gas phase or pure liquid phase.

**3. Analysis and Result 1: Otto cycle**

The thermodynamic Otto cycle, which is used as a model of spark-ignition (SI) internal combustion engine, is now discussed within the vdW fluid model. As depicted in Figs. 2(a) and 2(b), a single Otto cycle consists of four processes, i.e., two isentropic processes (compression stroke and expansion stroke) and two isochoric processes (combustion process and heat rejection). Here we intend to derive an exact expression of its engine efficiency.

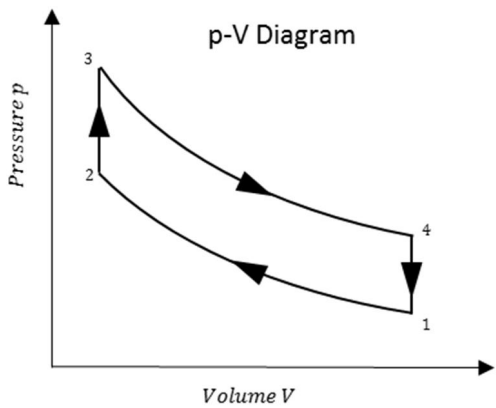

*Figure 2(a). p-V plot for Otto Cycle.*

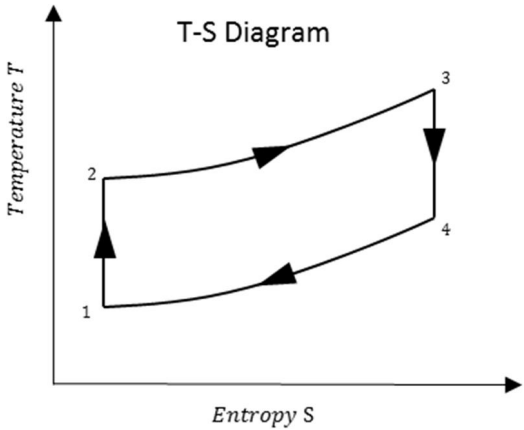

*Figure 2(b). T-S plot for Otto Cycle.*

First, the isentropic compression process denoted by 1→2 in each of the above figures is under consideration. By substituting Eqs. (1) and (2) into the First Law of thermodynamics

$$dQ = dU + pdV , \tag{7}$$

we can easily obtain

$$\frac{dQ}{T} = \frac{n\,c_v}{T} dT + \left(\frac{nR}{V-nb}\right) dV , \tag{8}$$

where the symbol $n$ denotes an average mole number of the vdW fluid available in the cycle. Requiring $dQ/T$ on the left-hand side here to vanish, and then substituting Eq. (6) for the heat capacity (of the liquid), followed by integrating from the starting point 1 to the endpoint 2, we can finally derive the relation

$$\frac{T_2+T_0}{T_1+T_0} = \left(\frac{v_1-b}{v_2-b}\right)^{2/d} , \tag{9}$$

expressed in terms of molar volumes $v_1$ and $v_2$. Similarly, in the isentropic expansion process denoted by 3→4, we can obtain

$$\frac{T_4+T_0}{T_3+T_0} = \left(\frac{v_3-b}{v_4-b}\right)^{2/d} \tag{10}$$

with molar volumes $v_3$ and $v_4$.

Next in the isochoric combustion process denoted by 2→3, we can derive an expression of the input heat by integrating Eq. (7) from the starting point 2 to the endpoint 3 with the help of Eqs. (1), (2) and (6) such that

$$Q_h = \frac{d}{2}nR\left\{T_3 - T_2 + T_0 \cdot \ln\left(\frac{T_2+T_0}{T_3+T_0}\right)\right\} . \tag{11}$$

Similarly, in the isochoric heat rejection process denoted by 4→1, we can obtain the expression of output heat

$$Q_c = \frac{d}{2}nR\left\{T_4 - T_1 + T_0 \cdot \ln\left(\frac{T_1+T_0}{T_4+T_0}\right)\right\} . \tag{12}$$

For later purposes, we remind here that

$$v_3 = v_2 , \tag{13}$$

$$v_4 = v_1 . \tag{14}$$

To compare this cycle later with the Carnot cycle, we now take the lowest cycle temperature $T_1$ as $T_c$, and the highest cycle temperature $T_3$ as $T_h$. Then, with the help of Eqs. (9)-(10) and (13)-(14), we can easily arrive at the expressions

$$\frac{T_2+T_0}{T_h+T_0} = \frac{T_c+T_0}{T_4+T_0} \tag{15}$$

$$T_2 + T_0 = \left(\frac{v_1-b}{v_2-b}\right)^{2/d}(T_c + T_0) , \tag{16}$$

which will be used below. Besides, the three dimensionless quantities will be useful, too, i.e., the cycle temperature ratio $t = T_h/T_c > 1$, the compression ratio $r = v_1/v_2 > 1$, as well as the volume ratio $0 < \bar{b} = b/v_2 < 1$. From this, a modified compression ratio may also be introduced for a later purpose

$$r' = \frac{r-\bar{b}}{1-\bar{b}} > 1 . \tag{17}$$

The efficiency of cycle is now ready to be considered,

$$\eta = 1 - \frac{Q_c}{Q_h}. \tag{18}$$

By substituting Eqs. (11) and (12) into this and utilizing Eqs. (15)-(17), we can, after some algebraic manipulations, acquire the Otto cycle efficiency for the vdW liquid,

$$\eta_O^{(l)} = 1 - \left(\frac{1}{r'}\right)^{2/d} \Xi_O^{(l)}, \tag{19}$$

where the expression

$$\Xi_O^{(l)} = \frac{1 - \lambda' + (r')^{2/d} \varphi \ln(\lambda')}{1 - \lambda' + \varphi \ln(\lambda')} \tag{20}$$

with

$$0 < \varphi = \frac{T_0}{T_h + T_0} < 1 \tag{21}$$

and

$$\lambda' = (r')^{2/d} \left\{\varphi + \frac{1}{t}(1 - \varphi)\right\}. \tag{22}$$

It is easy to see that with $T_0 \to 0$, the parameter $\varphi \to 0$, and so $\Xi_O^{(l)} \to \Xi_O^{(g)} = 1$, which immediately makes Eq. (19) reduce to the efficiency for the vdW gas,

$$\eta_O^{(g)} = 1 - \left(\frac{1}{r'}\right)^{2/d}. \tag{23}$$

Subsequently by setting $b=0$, the modified compression ratio $r' \to r$ in (17) indeed, from which Eq. (23) easily reduces to the well-known expression of Otto efficiency for the ideal gas [6,12],

$$\eta_O^{(id)} = 1 - 1/r^{2/d}. \tag{24}$$

Next we expand $\Xi_O^{(l)}$ of (20) at $\varphi = 0$ (i.e., $T_0 = 0$) to compare the magnitude of $\eta_O^{(l)}$ with that of $\eta_O^{(g)}$. After some algebraic manipulations, we can finally obtain from (19), for $\varphi \ll 1$,

$$\eta_O^{(l)} = \eta_O^{(g)} \{1 - f_O(x) \varphi\} + O(\varphi^2), \tag{25}$$

where

$$x = \lambda'(\varphi=0) = \frac{(r')^{2/d}}{t} \tag{26}$$

and

$$f_O(x) = \frac{\ln x}{1-x}. \tag{27}$$

It is then straightforward to verify algebraically that $f_O(x) < 0$, which will immediately yield that $\eta_O^{(l)} > \eta_O^{(g)}$ indeed (see the numerical analysis below, too). Further, by applying the expansion, valid for $\bar{b} \ll 1$, given by

$$\frac{1}{r'} = \frac{1}{r}\left\{1 + \left(\frac{1}{r} - 1\right)\bar{b}\right\} + O(\bar{b}^2), \tag{28}$$

we can obtain from (23)

$$\eta_O^{(g)} = \eta_O^{(id)}\left\{1 + \frac{(\cdots)}{\eta_O^{(id)}}\right\} + O(\bar{b}^2), \tag{29}$$

where

$$(\cdots) = \frac{2}{d}\left(\frac{1}{r}\right)^{2/d}\left(1 - \frac{1}{r}\right)\bar{b} > 0. \tag{30}$$

From this, we see that $\eta_O^{(g)} > \eta_O^{(id)}$ indeed. Along the same lines, the ratio $\eta_O^{(l)}/\eta_O^{(id)} > 1$ in the regime of both $\varphi \ll 1$ and $\bar{b} \ll 1$ will immediately appear from Eqs. (25) and (29) with the help of (26) and (28).

Now we numerically analyze the vdW efficiency expressions given in Eqs. (19) and (23). The figures below demonstrate their various behaviors, which are also in comparison with the ideal-gas counterpart in (24) and the Carnot efficiency $\eta_C$, altogether for a *fixed* temperature ratio *t*. This means that all cycles in consideration are assumed to operate between the same temperature extremes $T_h$ and $T_c$. We first observe that for a given compression ratio, the two vdW efficiencies, $\eta_O^{(l)}$ and $\eta_O^{(g)}$, are higher than their ideal-gas counterpart, respectively, as expected. Second, the former efficiency for the pure liquid phase is verified to be even higher than the latter one for the pure gas phase, especially in the low-temperature regime (i.e., in terms of the highest cycle temperature $T_h$). Third, because these efficiencies cannot go beyond the Carnot upper bound $\eta_C$ reflecting the Second Law, the physically allowed compression ratio $r$ is limited to $1 < r < r_L < r_G < r_I$, as demonstrated in Fig. 3, where the upper bounds $r_L$, $r_G$, and $r_I$ are given for the vdW liquid, the vdW gas, and the ideal gas, respectively. In fact, the three upper bounds can readily be determined by equating Eqs. (19), (23) and (24) to $\eta_C$, respectively. This means that if $r \geq r_{L/G/I}$, then the input heat can be shown to be $Q_h \leq 0$, which is however physically not allowed; e.g., in the ideal-gas model for simplicity, the efficiency can easily be rewritten as $\eta_O^{(id)} = 1 - T_1/T_2$ in terms of the cycle temperatures [6], while $\eta_C = 1 - T_1/T_3 > \eta_O^{(id)}$. If $\eta_O^{(id)} > \eta_C$, then it would lead to $T_2 \geq T_3$, which exactly means $Q_h \leq 0$. Our result demonstrates that the higher the efficiency is, the more shrinkage in the range of (physically allowed) compression ratio is found. Further, the efficiency $\eta_O^{(l)}$ itself tends to be higher in the regime of lower temperature, obviously with a more shrinkage in the compression ratio range.

Comments deserve here. First, comparison between the two cases of $\bar{b} = 0.2$ and $0.6$ is made in, e.g, Fig. 4 (obviously, $\bar{b} = 0$ for the ideal gas). As known, the molar parameters *b* are experimentally evaluated as $O(10^{-5})$ m$^3$/mol for typical elements [2]. Correspondingly, the molar volumes $v_2 \sim 5 \cdot 10^{-5}$ and $1.67 \cdot 10^{-5}$ m$^3$ are given for our numerical analysis, which are significantly small in magnitude. Those values of $\bar{b}$, with $V_2/n = v_2$, may be achieved, either (I) by taking a mole number $n$ significantly large, or (II) by taking a significantly compressed volume $V_2$, especially in the low-temperature regime (and so in the pure liquid phase). As demonstrated in Fig. 4, a smaller engine, i.e., with a larger value of $\bar{b}$ by following the scheme (II) for a given $b$, can then achieve higher efficiency. On the other hand, the engine efficiency should be independent of the mole number $n$ [cf. Eq. (18) with (11) and (12)].

Second, as discussed after Eq. (5), a vdW liquid, with its engine efficiency (19) with the help of (6), should meet the volume ratio $v_L/b$ condition given by the inequality with its upper bound of $1 + \{RbT/(2a)\}^{1/2}$. Accordingly, this upper bound is required to be sufficiently large (say, by the sufficiently small ratio $a/b$) that the maximum volume ratio $(v_L)_1/b$ of a cycle should be even smaller than this upper bound value. Third, our numerical analysis then implies that, considerably remarkably, a non-ideal fluid may provide better behaviors in engine performance (efficiency) than the standard ideal-gas model, if we can control the reduced range of the compression ratio.

Fourth, we briefly compare this theoretical model of SI engines with the same type of real engines. The phenomenon of knock is one of the major limitations for real SI engines, which occurs with an increase of compression ratio [11]. This results primarily from the autoignitions in the unburned part of the gas in the cylinder, which will lead to local, very high pressure peaks, thus leading to destroying the rim of the piston and other parts in the cylinder. Consequently, the compression ratio cannot be arbitrarily larger for a higher efficiency of real engine. In our model, likewise, an increase in magnitude of parameter $b$ as a measure of non-ideality renders effective volume $v - b$ smaller and then pressure $p$ larger [cf. (1)], which will, as discussed already, yield a shrinkage in the maximally allowed compression ratio, i.e., the impossibility of an arbitrary increase in compression ratio, for a *fixed* temperature ratio $t$ as a qualitative measure of engine efficiency. This means that to reveal the same performance, a vdW engine should be smaller than its ideal-gas counterpart. Therefore, this model may consistently reflect the real engine behaviors. This aspect is also true for Diesel engines to be in detail discussed in the following section. More comparison with real engines will be available as well, after the analysis of Diesel cycle.

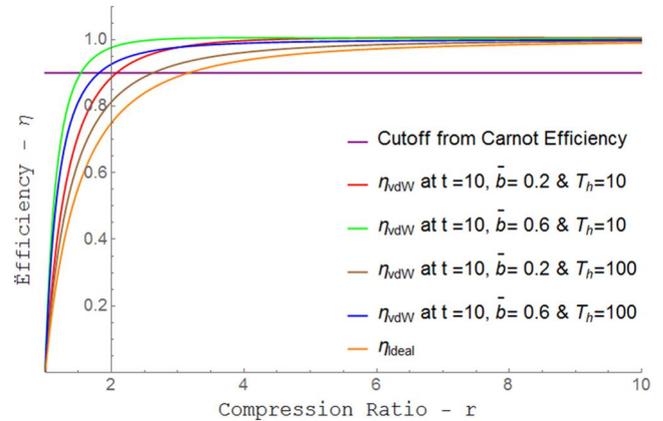

denotes Eq. (19), derived with the help of (6), valid in the region given by $1 < r \leq r_L$; the green dashed curve is given for (23) within $1 < r \leq r_G$; and the orange solid curve is for the ideal-gas case (24) within $1 < r \leq r_I$. In addition, the purple straight line is given for the maximum Carnot efficiency.

*Figure 4.* (Color online) *The Otto-cycle efficiency* (19) *in different cases for the monatomic gas with* $d=1$, *and* ($t=10$, $T_0=1$). *From top to bottom (at* $r=2$): *The green curve is given for* ($\bar{b}=0.6$, $T_h=10$); *the blue curve for* ($\bar{b}=0.6$, $T_h=100$); *the red curve for* ($\bar{b}=0.2$, $T_h=10$); *the brown curve for* ($\bar{b}=0.2$, $T_h=100$); *and the orange curve for the ideal gas. All curves are physically allowed in the respective regions only, where their values are not beyond the maximum Carnot efficiency* $\eta_C$ *given by the purple straight line.*

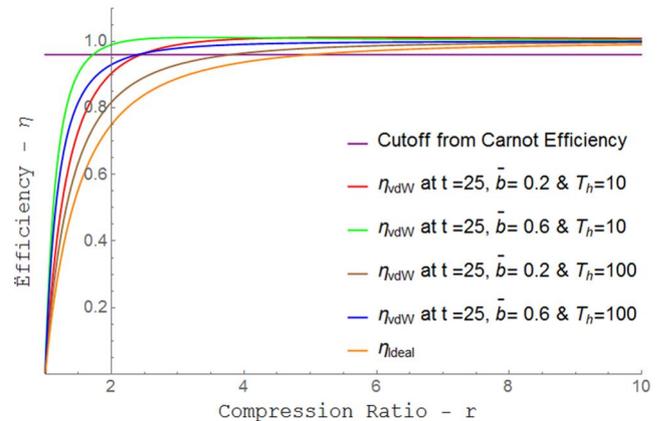

*Figure 5.* (Color online) *The same as Fig. 4, but for* $t=25$. *The efficiency values are higher than their counterparts of Fig. 4, respectively. The values observed beyond the Carnot value are physically not allowed.*

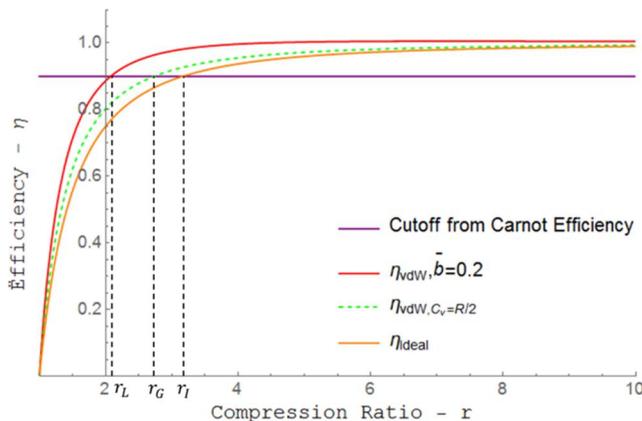

*Figure 3.* (Color online) *Otto-cycle efficiencies for the monatomic gas with* $d = 1$, *and* ($\bar{b} = 0.2$, $t = 10$, $T_h = 10$, $T_0 = 1$). *From top to bottom: The red solid curve*

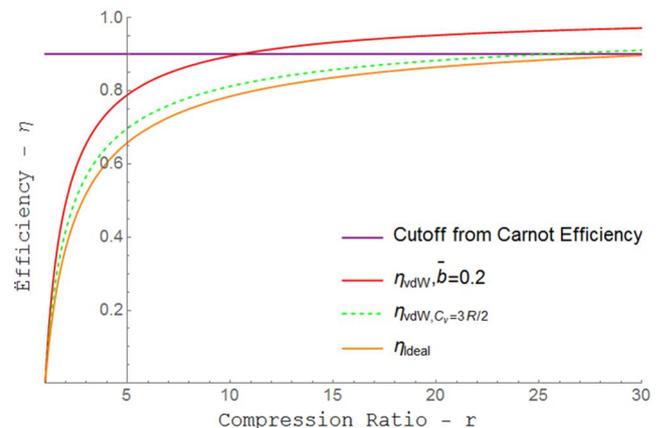

*Figure 6.* (Color online) *The same as Fig. 3, but for the monatomic gas in 3 dimensions* ($d=3$). *The ranges of*

*physically allowed compression ratios (determined from* $\eta < \eta_C$*) are enlarged, respectively, in comparison with those given in Fig. 3.*

## 4. Analysis and Result 2: Diesel cycle

The thermodynamic Diesel cycle, which is used as a model of compression-ignition (CI) internal combustion engine, is now discussed within the vdW fluid model. As depicted in Figs. 7(a) and 7(b), a single Diesel cycle consists of two isentropic processes and an isobaric process as well as an isochoric one. We will apply Eq. (6) for the heat capacity.

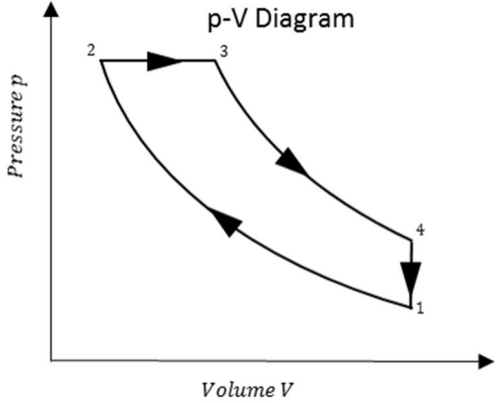

*Figure 7(a). p-V plot for Diesel Cycle.*

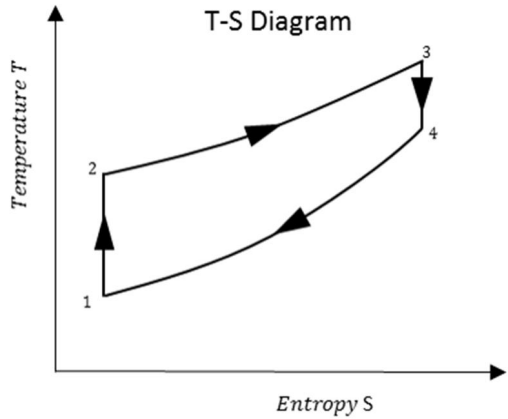

*Figure 7(b). T-S plot for Diesel Cycle.*

By construction, the isentropic compression process denoted by 1→2 and the isentropic expansion process denoted by 3→4 are identical to those of the Otto cycle, respectively. Therefore, Eqs. (9) and (10) are valid here, too. In the isobaric expansion process denoted by 2→3, heat is injected and its amount can be evaluated by integrating Eq. (7) from the starting point 2 to the endpoint 3, with the help of (1), (2) and (6), such that

$$Q_h = \int_2^3 \frac{d}{2} \frac{nR}{(T+T_0)} dT + \int_2^3 \left(p + \frac{n^2 a}{V^2}\right) dV ,\quad (31)$$

where the lower bound temperature $T_2$ and the lower bound volume $V_2$ are denoted, for simplicity, by 2, as well as the upper bounds $T_3$ and $V_3$ are denoted by 3, respectively; the symbol $n$ is a mole number. Due to the pressure being unchanged in this process, Eq. (31) can easily be evaluated as the expression in terms of the molar volumes,

$$Q_h = \frac{d}{2} nR \left(T_3 - T_2 + T_0 ln \frac{T_2 + T_0}{T_3 + T_0}\right) + n(v_3 - v_2)\left(\frac{RT_3}{v_3 - b} - \frac{a}{v_3^2} + \frac{a}{v_2 v_3}\right). \quad (32)$$

In the process denoted by 4→1, heat is rejected, and its amount is easily shown to be the same as Eq. (12) given for the Otto cycle.

Here we remind that

$$v_4 = v_1 . \quad (33)$$

We now take the lowest cycle temperature $T_1$ as $T_c$ and the highest cycle temperature $T_3$ as $T_h$. With the help of (9)-(10) and (33), the temperatures $T_2$ and $T_4$ can then be expressed as

$$T_2 = \left(\frac{v_1 - b}{v_2 - b}\right)^{2/d} (T_c + T_0) - T_0 \quad (34)$$

$$T_4 = \left(\frac{v_3 - b}{v_1 - b}\right)^{2/d} (T_h + T_0) - T_0 , \quad (35)$$

which will easily be transformed into

$$T_2 + T_0 = (T_c + T_0)\left(\frac{v_1 - b}{v_2 - b}\right)^{2/d} \quad (36)$$

$$\frac{1}{T_4 + T_0} = \frac{1}{T_h + T_0}\left(\frac{v_1 - b}{v_3 - b}\right)^{2/d} , \quad (37)$$

respectively, to be used below. Besides, we now introduce the cut-off ratio $r_c = v_3/v_2 < r$. From this, a modified cut-off ratio can also be introduced for a later purpose,

$$r_c' = \frac{r_c - \bar{b}}{1 - \bar{b}} < r' . \quad (38)$$

Also, from the isobaric process 2→3 ($p_2 = p_3$), we can easily find, with the help of (1), that

$$\frac{T_2}{T_3} = \frac{a(v_2 - b)}{T_h R}\left(\frac{1}{v_2^2} - \frac{1}{v_3^2}\right) + \frac{v_2 - b}{v_3 - b} . \quad (39)$$

Dividing Eq. (34) by $T_3$ (i.e., $T_h$) and equating it with Eq. (39), we can then obtain another expression of temperature ratio $t = T_h/T_c$, given by

$$\frac{1}{t} = \frac{\bar{a}(1-\bar{b})\left(1 - \frac{1}{r_c^2}\right) + \frac{1}{r_c'} + \frac{T_0}{T_h}}{(r')^{2/d}} - \frac{T_0}{T_h}, \quad (40)$$

where the dimensionless quantity $\bar{a} = a/(v_2 R T_h)$. With the help of (21), Eq. (40) can easily be rewritten as

$$\frac{1}{t} = \frac{\varphi}{1-\varphi}\left\{\left(\frac{1}{r'}\right)^{2/d} - 1\right\} + \left(\frac{1}{r'}\right)^{2/d}\left\{\bar{a}(1 - \bar{b})\left(1 - \frac{1}{r_c^2}\right) + \frac{1}{r_c'}\right\}, \quad (41)$$

which will be useful below. Its behaviors are demonstrated in the figures below; by construction, it is true that $0 < \varphi < 1$, and $0 < 1/t < 1$.

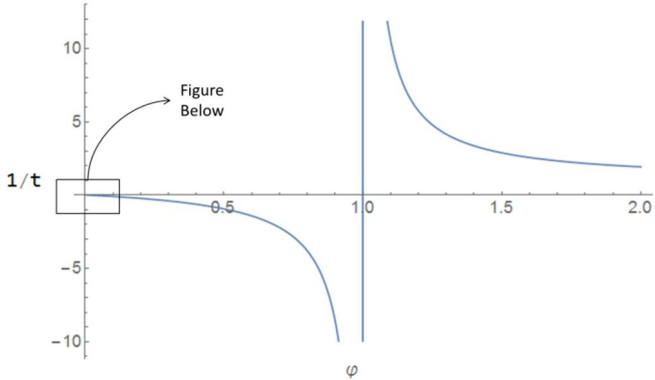

*Figure 8(a).* (Color online) $1/t$ versus $\varphi$ for $(d = 1, \bar{a} = 0.02, R = 1, T_h = 10, \bar{b} = 0.2, r = 4, r_c = 2)$.

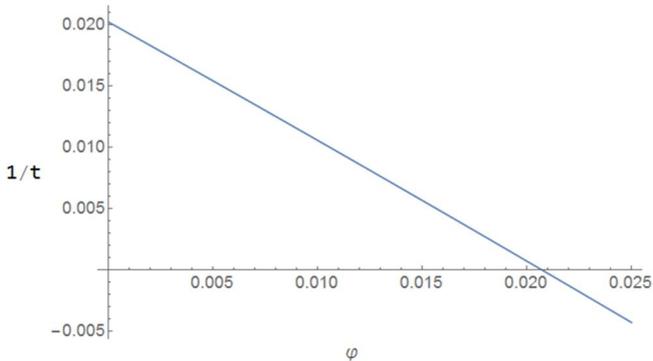

*Figure 8(b).* (Color online) *An enlarged plot of the inset given in Fig. 8(a)*: $0 < 1/t = 0.976 - 0.956/(1-\varphi) < 1$. If $\varphi \to 0.021$, then $1/t \to 0$.

Now we evaluate the Diesel engine efficiency. By substituting Eqs. (12), (32) and (36)-(37) into (18), we can, after some algebraic manipulations, arrive at its expression for the vdW liquid, given by

$$\eta_D^{(l)} = 1 - \left(\frac{r_c'}{r'}\right)^{2/d} \Xi_D^{(l)}, \quad (42)$$

where the expression

$$\Xi_D^{(l)} = \frac{1 - \lambda_D' + (r'/r_c')^{2/d} \, \varphi \, \ln(\lambda_D')}{1 - \lambda' + \varphi \, \ln(\lambda') + \frac{2}{d}\left\{1 - \frac{1}{r_c'} + \bar{a}\left(1 - \frac{1}{r_c}\right)^2\right\}} \quad (43)$$

with $\lambda_D' = \lambda'/(r_c')^{2/d}$. It is straightforward to see that with $T_0 \to 0$ (i.e., $\varphi \to 0$), Eq. (43) reduces to

$$\Xi_D^{(g)} = \frac{1 - \lambda_D'}{1 - \lambda' + \frac{2}{d}\left\{1 - \frac{1}{r_c'} + \bar{a}\left(1 - \frac{1}{r_c}\right)^2\right\}}, \quad (44)$$

leading to the Diesel efficiency for the vdW gas

$$\eta_D^{(g)} = 1 - \left(\frac{r_c'}{r'}\right)^{2/d} \Xi_D^{(g)}. \quad (45)$$

Further, by letting $(\bar{a}, \bar{b}) \to 0$, equivalently, $r' \to r$ and $r_c' \to r_c$, so leading to $\lambda' \to 1/r_c$ and $\lambda_D' \to 1/r_c^{1+(2/d)}$ with the help of $t = r_c \, r^{2/d}$ [cf. (41)], Eq. (44) gives the well-known expression for the ideal-gas model

$$\eta_D^{(id)} = 1 - \frac{\{r_c^{1+(2/d)} - 1\} d}{r^{2/d} \, (r_c-1)(d+2)}. \quad (46)$$

Comments deserve here. First, it is easy to see that with the cut-off ratio $r_c \to 1$, Eqs. (42), (45) and (46) reduce to their Otto engine counterparts (19), (23) and (24), respectively. Second, it will be useful later to solve Eq. (41) for $r_c = r_c(r, \bar{a}, \bar{b}, T_h, T_0, t)$, i.e., especially with a given ratio $t$. Substituting this into (42), (45) and (46) will then lead to the respective expressions of efficiency, without $r_c$. From Fig. 7(b), we can then observe that for $(\bar{a}, \bar{b}, T_h, T_0)$ and the temperature ratio $t$ kept fixed, an increase of compression ratio $r$ (leading to a larger amount of temperature increase during process 1→2) must correspondingly give rise to a decrease of input heat $Q_h$ (during process 2→3) and so a decrease of $r_c$ [cf. (31)]. This can easily be seen, especially in the ideal gas model, from $r_c = t/r^{2/d}$, which will also transform (46) into

$$\eta_D^{(id)} = 1 - \frac{\{t/r^{(\gamma-1)}\}^\gamma - 1}{\gamma \, t \, \{1 - r^{(\gamma-1)}/t\}}. \quad (47)$$

Here the symbol $\gamma = c_p/c_v = 1 + (2/d)$ denotes the ratio of molar heat capacity at constant pressure to molar capacity at constant volume. Eq. (47) has already been derived in, e.g., [12] where comparison of various model engine cycles has been carried out for the same temperature ratio $t$, but within the ideal gas only. Here we can also verify, from $r_c > 1$, i.e., $t > r^{2/d}$, that the Carnot efficiency $\eta_C$ is greater indeed than the Otto efficiency $\eta_O^{(id)}$ being, in turn, greater than the Diesel efficiency $\eta_D^{(id)}$, easily shown from (46).

Next, we expand $\Xi_D^{(l)}$ of (43) at $\varphi = 0$, as we did for the Otto cycle. Then, it is straightforward to obtain

$$\left.\frac{d\lambda'}{d\varphi}\right)_{\varphi=0} = 1 - \bar{a}(1 - \bar{b})\left(1 - \frac{1}{r_c^2}\right) - \frac{1}{r_c'} \quad (48)$$

and

$$\left.\frac{d\lambda_D'}{d\varphi}\right)_{\varphi=0} = \left(\frac{1}{r_c'}\right)^{2/d} \left.\frac{d\lambda'}{d\varphi}\right)_{\varphi=0}. \quad (49)$$

Also, with the help of (41), Eq. (26) reduces to

$$x = \lambda'(\varphi=0) = \bar{a}(1 - \bar{b})\left(1 - \frac{1}{r_c^2}\right) + \frac{1}{r_c'}. \quad (50)$$

Then, we can finally find that for $\varphi \ll 1$,

$$\eta_D^{(l)} = \eta_D^{(g)} \{1 - f_D(x) \, \varphi\} + \mathcal{O}(\varphi^2),$$

where

$$f_D(x) = \frac{\ln x - \frac{\{2/\eta_D{}^{(g)}\}}{d}\ln(r_c') + \left\{\frac{\eta_O{}^{(g)}}{\eta_D{}^{(g)}} - 1\right\}(1-x)}{1 - x + \frac{2}{d}\left\{1 - \frac{1}{r_c} + \bar{a}\left(1 - \frac{1}{r_c}\right)^2\right\}}. \quad (52)$$

Fig. 9 demonstrates that $f_D(x) < 0$, which will yield that $\eta_D{}^{(l)} > \eta_D{}^{(g)}$ indeed

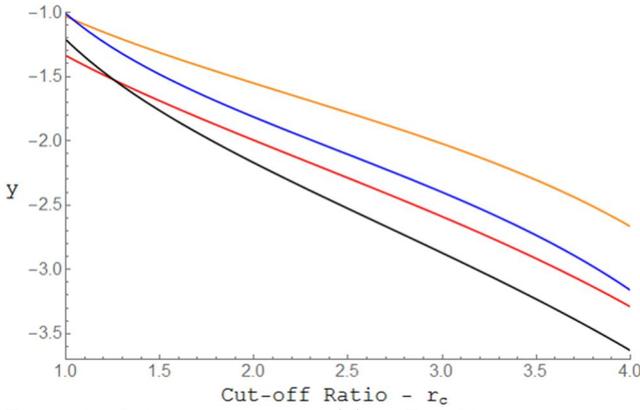

*Figure 9.* (Color online) $y = f_D(x)$ of Eq. (52) versus the cut-off ratio $r_c$ with $r = 4$ and $\bar{a} = 0.02$. From top to bottom (at $r_c = 4$): The orange curve with $d = 1$ and $\bar{b} = 0.2$; the blue curve with $d = 1$ and $\bar{b} = 0.6$; the red curve with $d = 5$ and $\bar{b} = 0.2$; and the black curve with $d = 5$ and $\bar{b} = 0.6$.

Now we numerically analyze the expressions (42), (45) and (46) of the vdW Diesel efficiency. The figures below provide their behaviors, which are qualitatively the same as those of the Otto efficiency expressions (19), (23) and (24), respectively. These behaviors will also justify for Diesel engines the same remark upon comparison between the vdW model engine and the corresponding real one as that given for Otto engines before Fig. 3. Here, an amount of input heat $Q_h > 0$, consisting with $r_c > 1$. As demonstrated in Fig. 12, the Diesel efficiency in (42) is smaller than the Otto efficiency in (19) for the same compression ratio, as in the case of ideal gas model.

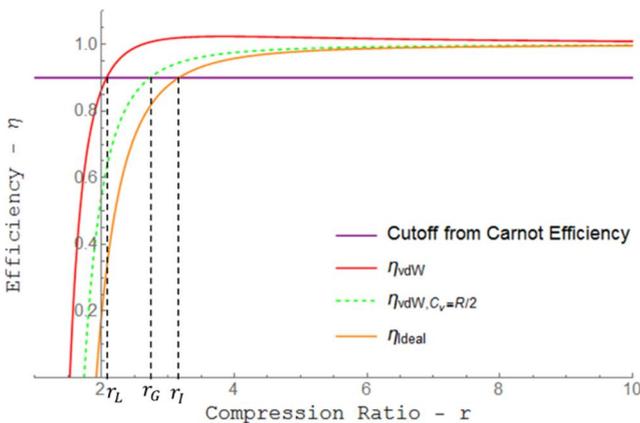

*Figure 10.* (Color online) Diesel-cycle efficiencies for the monatomic gas with $d = 1$, and ($T_h = 10, T_0 = 1, \bar{a} = 0.001, \bar{b} = 0.2$), as well as a given ratio $t = 10$. From top to bottom (at $r = 2$): The red solid curve denotes Eq. (42), derived with the help of (6), being valid in the region given by $1 < r \leq r_L$; the green dashed curve is given for (45) within $1 < r \leq r_G$; and the orange solid curve is for the ideal-gas case (46) within $1 < r \leq r_I$. In addition, the purple straight line is given for the maximum Carnot efficiency $\eta_C$. The values observed beyond the Carnot value are physically now allowed.

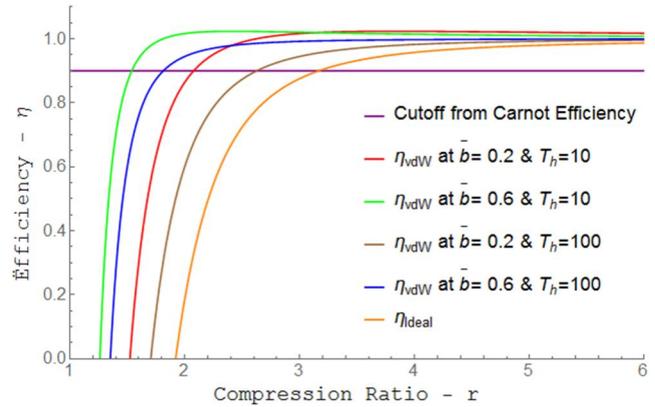

*Figure 11.* (Color online) The Diesel-cycle efficiency (42) in different cases for the monatomic gas with $d=1$, and $\{T_0 = 1, t = 10, a/(v_2 R) = 0.01\}$. From the leftmost point of intersection with $\eta = \eta_C$ to the rightmost: The green curve is given for ($\bar{b}=0.6$, $T_h=10$) and so $\bar{a} = 0.001$; the blue curve for ($\bar{b}=0.6$, $T_h=100$) and so $\bar{a} = 0.0001$; the red curve for ($\bar{b}=0.2$, $T_h=10$) and so $\bar{a} = 0.001$; the brown curve for ($\bar{b}=0.2$, $T_h=100$) and so $\bar{a} = 0.0001$; and the orange curve for the ideal gas. All curves are physically allowed in the respective regions only, where their values are not beyond the maximum Carnot efficiency given by the purple straight line. Also, the compression ratio $r$ must be greater than the cut-off ratio $r_c$ given by the x-axis intersection point.

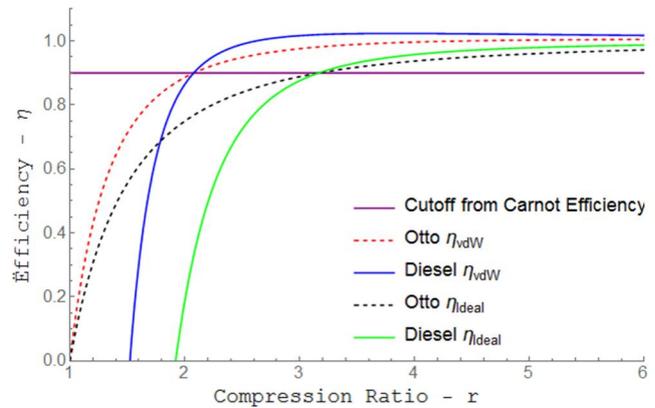

*Figure 12.* (Color online) Diesel-cycle efficiency versus Otto-cycle efficiency for the monatomic gas with $d=1$, and $(T_0 = 1, T_h = 10, t = 10, \bar{a} = 0.001, \bar{b} = 0.2)$. For the solid curves, from top to bottom: The blue one for Diesel efficiency (42); and the green one for its ideal-gas counterpart (46). For the dashed curves, from the top to bottom: The red one for Otto efficiency (19); and the black one for its ideal-gas counterpart (24). The values observed beyond the Carnot value are physically not allowed.

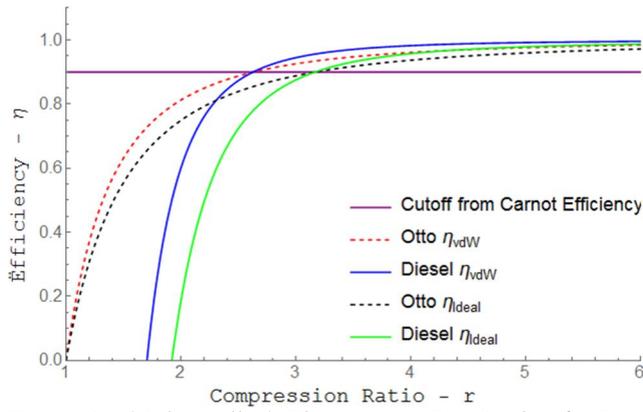

*Figure 13.* (Color online) *The same as Fig. 12, but for $T_h = 100$ (high-temperature regime), corresponding to the gas phase, while $T_h = 10$ (low-temperature regime) in Fig. 12 on the other hand, corresponding to the liquid phase relatively. Each efficiency is observed to be smaller, and valid in the larger range of compression ratio, than its counterpart given in Fig. 12.*

Comments on additional comparison with real engines are now provided. First, it is known that the highest temperature produced by a typical real internal combustion engine is lower (equivalent to our temperature ratio $t$ being smaller) than its counterpart calculated from the ideal-gas model [11]. In fact, some various non-ideal gas models have been applied to a Diesel cycle, to explain this experimental phenomenon [13]; it has then been shown that for given ratios $r$ and $r_c$, the vdW model can provide the closest fitting to such a real-gas behavior, simply by using the specific pair $(a_0, b_0)$ for air only. Now, Fig. 14 plots the behaviors of ratio $t$ in response to variations of $(\bar{a}, \bar{b})$, to reveal an influence of the vdW parameters explicitly; if $\bar{b}$ is extremely small (being the case for the fitting considered in [13]), then the reciprocal ratio $1/t$ still remain greater than its ideal-gas counterpart, exactly corresponding to its value of bottom curve at $\bar{b} = 0$ therein, and thus the ratio $t$ smaller, which will imply a lower efficiency of real engine. As demonstrated, however, if $\bar{b} = b/v_2$ continues to increase, say, by reducing the engine size $(v_2)$ gradually, then the ratio $t$ increases and finally goes beyond its ideal-gas counterpart, thus leading to a higher efficiency, consisting with our above results.

Second, the ignition delay in a Diesel engine is defined as the time interval between the start of injection and the start of combustion [11]. As known, increasing the ignition delay tends to reduce a tendency for knock. However, a closer contact between working-fluid molecules with finite size will result in their higher density and a higher internal pressure, subsequently reducing the ignition delay period. In our model analysis, consistently, an increase in magnitude of the vdW parameters enables the pressure to be higher than that resulting from the ideal-gas model; as discussed already, this scenario will then yield a shrinkage in the maximally allowed compression ratio $r$, i.e., the impossibility of an arbitrary increase in $r$, for a fixed temperature ratio $t$ as a qualitative measure of engine efficiency.

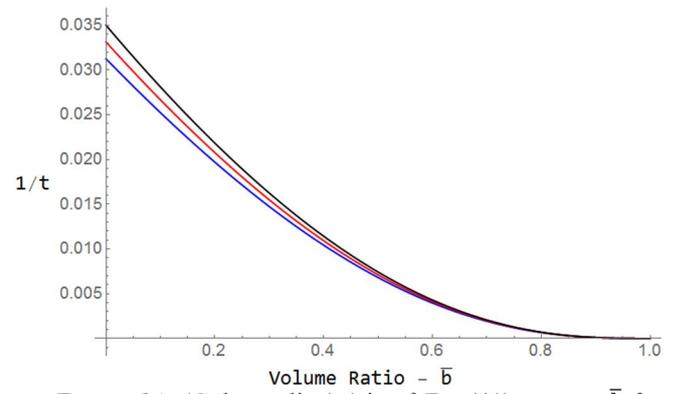

*Figure 14.* (Color online) $1/t$ of Eq. (41) versus $\bar{b}$ for (d = 1, $T_0 = 0, r = 4, r_c = 2$). *From bottom to top*: *The blue curve with $\bar{a} = 0$; the red curve with $\bar{a} = 0.04$; and the black curve with $\bar{a} = 0.08$. The case of $\bar{a} = \bar{b} = 0$ exactly corresponds to the ideal gas.*

## 5. Conclusion

In this paper, we studied the engine efficiencies of both Otto and Diesel cycles as the schematic models of internal combustion engines, by employing a working fluid of the van der Waals model supplemented by a temperature-dependent heat capacity (i.e., a *generalized* van der Waals fluid model conforming with the Third Law). We systematically compared our exact results with those of the corresponding ideal-gas model and the Carnot cycle, for a fixed ratio of maximum and minimum cycle temperatures. We also provided comparison between the van der Waals model engine and the real ones. We then found that it is possible for an engine running upon a non-ideal fluid model to achieve a higher efficiency than that obtained from the ideal-gas model. This "good" news is considerably interesting, as compared to the known "bad" news that the actual non-ideal gas in thermal processes of the typical internal combustion engines of industry would produce, for a given internal pressure, the *lower* internal temperature (as a measure of the available thermal exergy) than its counterpart calculated from the ideal gas model.

Our analysis implies that an appropriate selection of working fluid (i.e., in terms of the parameters *a* and *b* in the vdW model) should also be considered in the engine performance study, in addition to the engine architectures and bath temperatures. It will then contribute to providing a fundamental guidance for the efficient management of thermal and mechanical energy at macro level, say, in optimal design of next-generation heat engines in automobile industry.

## Acknowledgements
We acknowledge financial support provided by the U.S. Army Research Office (Grant No. W911NF-15-1-0145) and the Center for Energy Research and Technology, NC A&T.

## Nomenclature

| | |
|---|---|
| $p$ | Pressure, Pa |
| $v$ | Molar volume, m$^3$ mol$^{-1}$ |
| $R$ | Universal gas constant, J K$^{-1}$ mol$^{-1}$ |
| $T$ | Temperature, K |
| $U$ | Internal energy, J |
| $T_c$ | Cold bath temperature, K |
| $T_h$ | Hot bath temperature, K |
| $t$ | Temperature ratio ($T_h/T_c$) |

| $\eta_\text{C}$ | Carnot cycle efficiency |
| --- | --- |
| $\eta_\text{O}$ | Otto cycle efficiency |
| $\eta_\text{D}$ | Diesel cycle efficiency |
| vdW | van der Waals |
| $a$ | vdW parameter 1, Pa m$^6$ mol$^{-2}$ |
| $b$ | vdW parameter 2, m$^3$ mol$^{-1}$ |
| $C_V$ | Heat capacity at constant volume, J K$^{-1}$ |
| $c_v$ | Molar heat capacity at constant volume, J K$^{-1}$ mol$^{-1}$ |
| $c_p$ | Molar heat capacity at const. pressure, J K$^{-1}$ mol$^{-1}$ |
| $\gamma$ | Ratio $c_p/c_v$ |
| $v_c$ | Critical molar volume, m$^3$ mol$^{-1}$ |
| $p_c$ | Critical pressure, Pa |
| $T_{cr}$ | Critical temperature, K |
| $T_r$ | Reduced (dimensionless) temperature ($T/T_{cr}$) |
| $v_G$ | Molar volume in gas phase, m$^3$ mol$^{-1}$ |
| $v_{GL}$ | Molar volume in gas-liquid two phase, m$^3$ mol$^{-1}$ |
| $v_L$ | Molar volume in liquid phase, m$^3$ mol$^{-1}$ |
| $T_0$ | Constant term introduced in heat capacity model, K |
| $n$ | Number of moles |
| $d$ | Number of degrees of freedom |
| $Q_h$ | Input heat, J |
| $Q_c$ | Output heat, J |
| $\bar{a}$ | Dimensionless ratio $\{a/(v_2 R T_h)\}$ |
| $\bar{b}$ | Dimensionless ratio $(b/v_2)$ |
| $r$ | Compression ratio |
| $r_c$ | Cut-off ratio |